# *Fermi*-LAT Observation of Quiescent Solar Emission

Elena Orlando
*Max-Planck-Institut für extraterrestrische Physik, Postfach 1312, 85741 Garching, Germany*
Nicola Giglietto
*INFN Bari and Dipartimento Interateneo di Fisica- Universita' e Politecnico di Bari*
On behalf of the Fermi Large Area Telescope Collaboration

The Large Area Telescope (LAT) on board Fermi has detected high-energy gamma rays from the quiet Sun produced by interactions of cosmic-ray nucleons with the solar surface and cosmic-ray electrons with solar photons in the heliosphere. Such observations provide a probe of the extreme conditions near the solar atmosphere and photosphere and permit the study of the modulation of cosmic rays over the inner heliosphere. For the first year of Fermi observations the solar modulation was at its minimum corresponding to a maximum cosmic-ray flux and, hence, maximum gamma-ray emission from the Sun. We discuss the study of the quiescent solar emission, including spectral analysis of its two components, disk and inverse Compton, using the first-year data of the mission and models using the electron spectrum measured by Fermi.

## 1. INTRODUCTION

*Fermi* was successfully launched from Cape Canaveral on the 11th of June 2008. It is currently in a circular orbit around the Earth at an altitude of 565 km having an inclination of 25.6° and an orbital period of 96 minutes. After an initial period of engineering data taking, the observatory was put into a sky-survey mode. The observatory has two instruments onboard, the Large Area Telescope (LAT) [1] pair-conversion gamma-ray detector and tracker and a Gamma Ray Burst Monitor (GBM), dedicated to the detection of gamma-ray bursts. The LAT instrument on *Fermi* provides coverage over the energy range from 30 MeV to several hundred of GeV.

The solar disk is a steady source of gamma rays produced by hadronic interactions of cosmic rays (CRs) in the solar solar atmosphere and photosphere, the disk emission [2]. The estimated disk flux of $\sim 10^{-7}$ cm$^{-2}$s$^{-1}$ above 100 MeV from pion decays was at the limit of EGRET sensitivity [3]. Moreover the Sun was predicted to be an extended source of gamma-ray emission, produced by inverse Compton (IC) scattering of cosmic-ray electrons off solar photons in [4] and [5]. This emission has a broad distribution on the sky with maximum intensity in the direction of the Sun. A detailed analysis of the EGRET data [6] yielded a total flux of $(4.4\pm2.0) \times 10^{-7}$ cm$^{-2}$s$^{-1}$ within a region of radius 20° around the Sun for E >100 MeV for these two solar components, consistent with the predicted level.

Both emission mechanisms have their maximum flux at solar minimum condition when the cosmic-ray flux is maximum because of the lower level of solar modulation.

Observations with *Fermi*-LAT of the inverse Compton scattered solar photons allow for continuous monitoring of the cosmic-ray electron spectrum potentially even in the close proximity of the solar surface.

We report on the first months of observations and preliminary estimates of observed fluxes and emission profiles.

## 2. DATA SELECTION

During the first year of the *Fermi*-LAT mission, the Sun was at its minimum activity in the beginning of Solar Cycle 24. The quiescent solar gamma-ray flux during this period is therefore expected to be at its maximum. LAT is able to detect the solar emission almost daily when the Sun is not close to the Galactic plane or bright sources. Figure 1 shows the count map above 100 MeV in Sun-centred coordinates, integrated for a period from August 2008 and July 2009. A significant photon excess corresponding to the position of the Sun is clearly visible. This image is an update of the count map published in [7] with data accumulated from July 2008 to the end of September 2008 and in [8] with data collected during the first 6 months of the mission, from August 2008 to January 2009.

A preliminary analysis of the solar emission, using the standard unbinned maximum likelihood LAT science tool was performed by fitting a model for the background and a point source with a simple power law for the Sun [9]. This was an approximate method since a more precise analysis should take into account the two different components of the solar emission, as described in the next Section. In this analysis we improved the fit by adding a template for the Inverse Compton emission. We then used the standard unbinned maximum likelihood LAT science.

In order to remove contamination from the expected lunar emission, we exclude the data when the angular separation of the Moon and Sun is less than 30°. Moreover, to avoid gamma rays emitted from the Earth's atmosphere we only accepted events at zenith angles less than 105°. For this analysis we use the "Diffuse" gamma-ray event class [1], corresponding to events with the highest probability of being photons. We also use Science Tools and IRFs (Instrumental Response Functions) version P6_V3 [11]. We exclude data when the brightest sources with a flux higher than $5 \times 10^{-7}$ cm$^2$s$^{-1}$ above 100 MeV, are within 5° degrees of the Sun. Similarly we exclude the Galactic plane. The solar minimum conditions and the absence of gamma-ray flares during the 1st year of the LAT mission are extremely favorable for observations of the quiet sun emission.

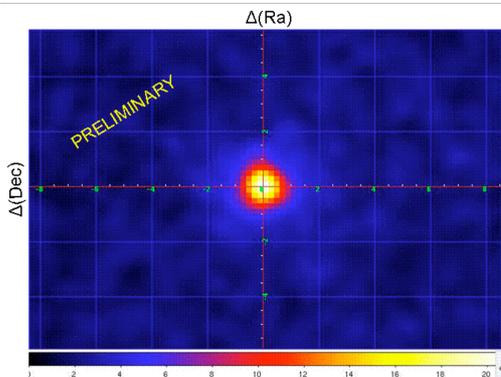

Figure 1: Count map in Sun-centred system for E>100MeV obtained with *Fermi*-LAT using data between August 2008 and July 2009. The pixel size used is 0.25° and the image covers 20°×10°. Counts per pixel are shown on the color bar. The image is 2-bin-radius Gaussian smoothed.

## 3. ANALYSIS METHOD

As the Sun is a moving source, we use software developed within the collaboration to analyze the data in a Sun-centred coordinate system.

We utilize a unique technique, the 'fake' source method, to remove the anisotropic background component that includes diffuse Galactic and extragalactic emission, the instrumental background and sources not excluded from the data. This method follows a source moving along the same ecliptic path as the Sun, but at a different time, with the same cuts applied to the solar data, in order to remove this residual background. We then have used the standard unbinned maximum likelihood provided with the LAT Science Tools. We performed a preliminary analysis using a background based on the 'fake Sun' data, a power law function to represent the disk emission, and a model for the inverse Compton component using the electron spectrum measured by Fermi [12].

## 4. RESULTS

We present here the preliminary results, angular profiles of counts and fluxes, obtained with the fitting technique explained in Section 3.

In Figure 2 we show that a combination of background and a solar disk component cannot explain the angular distribution of events and that an inverse Compton component is needed. The plots show the event density >500 MeV (black) up to 5 and 20 degrees from the Sun. Plotted for comparison are the background model and the sum of the background and fitted disk component. In fact the data confirm that the observed shape is not in agreement with a disk source only, for the calculated point spread function of the LAT, in contrast to the emission from the Moon [10], which is consistent with a point source.

We estimate that the overall solar emission has been detected at ~40 sigma level of confidence; however we still have signficant uncertainty in the determination of the relative contributions of the solar disk and inverse Compton components. We used a power law to represent the disk component and model for the inverse Compton component based on the measured electron spectrum >100 MeV at Earth. The flux of the disk component is ~3.2 $\times$ 10$^{-7}$ cm$^{-2}$s$^{-1}$, while that of IC component is ~1.5 $\times$ 10$^{-6}$ cm$^{-2}$s$^{-1}$ within a region of 20° radius centred on the Sun. These flux estimates still have an uncertainty of 50% primarily due to uncertainties in the IC model and the modulation of cosmic ray electrons and protons in the inner heliosphere. The uncertainty also includes systematic errors of about 20% for the instrument response

The resulting fluxes are in general agreement with theoretical models for both inverse Compton [4][5] and disk [2] emission and with the previous detection with EGRET [6].

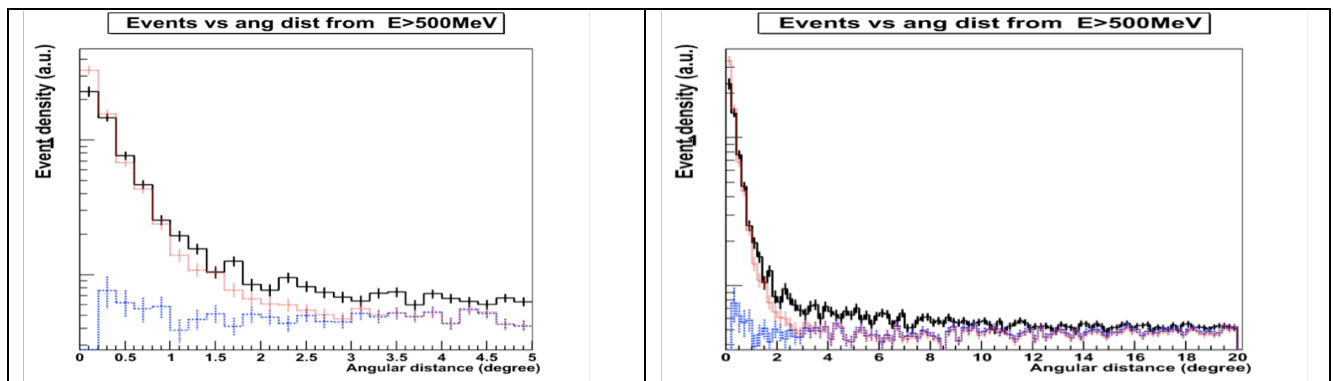

Figure 2: Event density profiles, in arbitrary units, with angular distance up to 5° (left) and 20° (right) from the Sun for energies above 500 MeV. Black lines are the observed counts, blue lines are the background and red lines are the simulation background plus the model disk component. It is clear that the disk component alone cannot explain the data and an extra extended component given by inverse Compton emission is needed as expected.

## 5. CONCLUSIONS

The quiet solar emission is detected with high statistical significance ($> 40\sigma$) and two components have been clearly separated for the first time.

Clear evidence of the detection of both components is shown in the density profiles plots. It is clear that a disk component alone cannot explain the data and an extra extended component given by inverse Compton emission is needed. Preliminary fluxes are given for both components, but the analysis made to obtain them is model dependent. Hence, a deeper analysis is being undertaken.

## 6. ACKNOWLEDGMENTS

The Fermi LAT Collaboration acknowledges support from a number of agencies and institutes for both development and the operation of the LAT as well as scientific data analysis. These include NASA and DOE in the United States, CEA/Irfu and IN2P3/CNRS in France, ASI and INFN in Italy, MEXT, KEK, and JAXA in Japan, and the K. A. Wallenberg Foundation, the Swedish Research Council and the National Space Board in Sweden. Additional support from INAF in Italy for science analysis during the operations phase is also gratefully acknowledged.